# Absence of Spin Hall Magnetoresistance in Pt/(CoNi)$_n$ multilayers

Yongwei Cui,[1] Xiaoyu Feng,[1] Qihan Zhang,[1] Hengan Zhou,[2] Wanjun Jiang,[2] Jiangwei Cao,[1] Desheng Xue,[1] and Xiaolong Fan[1*]

[1]*Key Laboratory for Magnetism and Magnetic Materials of the Ministry of Education, Lanzhou University, Lanzhou 730000, People's Republic of China*
[2]*State Key Laboratory of Low-Dimensional Quantum Physics and Department of Physics, Tsinghua University, Beijing 100084, China*

Correspondence and requests for materials should be addressed to:
*fanxiaolong@lzu.edu.cn

**Abstract**

We systematically studied the magnetoresistance effect in a Pt/(CoNi)$_n$ multilayer system with perpendicular magnetic anisotropy and the fcc (111) texture. The angular dependence of magnetoresistance, including high-order cosine terms, was observed in a plane perpendicular to the electrical current; this was attributed to the geometrical-size effects caused by crystal symmetry, the ordered arrangement of grains, and the anisotropic interface magnetoresistance effect caused by the breaking of the symmetry at interfaces. Based on the accuracy of our experimental results, the magnitude of spin Hall magnetoresistance (SMR) in Pt/(CoNi)$_n$ was expected to be below $1\times10^{-4}$. However, on evaluating the spin Hall angle of $\geq 0.07$ for Pt using spin-torque ferromagnetic resonance measurements, the theoretical magnitude of SMR in our samples was estimated to exceed $7\times10^{-4}$. This absence of SMR in the experimental results can be explained by the Elliott-Yafet spin relaxation of itinerant electrons in the ferromagnetic metal, which indicates that the boundary conditions of the spin current in the heavy metal/ferromagnetic insulator may not be applicable to all-metallic heterostructures.

## I. INTRODUCTION

Spin Hall magnetoresistance (SMR) refers to the anisotropic absorption of spin currents in heavy metals (HMs); this absorption is determined by the direction of magnetization in the adjacent ferromagnetic layer (FML) and represented by a change in the HM resistance along the direction of magnetization [1-3]. The SMR essentially reflects the interaction between spin currents and the local magnetic moment based on the conversion between spin and charge currents in a heterostructure; therefore, it is a powerful index for evaluating the spin-orbit coupling in heterogeneous structures. In particular, it can be used to determine the spin Hall angle, spin diffusion length in HMs, and interface spin mixing conductance [2-5]. Furthermore, interface spin-orbit coupling can be studied using spin-orbit MR (SOMR) [6]. In addition, SMR can be used to study the magnetization in magnetic insulators and monitor the directions of the Néel order in antiferromagnetic materials through highly sensitive electrical measurements [7-9].

Therefore, SMR plays a critical role in spintronics research.

A model for the SMR in the heterostructures of HMs and ferromagnetic insulators (FMI) has been proposed [2, 10, 11] for analyzing spin-dependent transport behaviors in all-metallic heterostructures [4, 12]. Compared with that in HM/FMI, the SMR effect in HM/ferromagnetic metals (FMs) is more complicated owing to the influence of interface spin-orbit coupling on conducting electrons and the conversion of charge and spin currents in FMs. Therefore, based on previous theoretical research, the absorption of the longitudinal spin current by the FM layer [4], spin current generated by the FM layer [13], and anomalous Hall effect of the FM layer [5, 14, 15] should be considered. More importantly, although the appearance of the magnetoresistance effect in the plane perpendicular to the charge current (i.e., the yz-plane) is a feature that distinguishes SMR from the traditional anisotropic magnetoresistance (AMR), the ordered arrangement of grains in FMs and the breaking of symmetry at interfaces also generates magnetoresistance in that plane [16-18]. For instance, Kobs et. al. observed a distinct magnetoresistance in the yz-plane for a Pt/Ni/Pt system; however, the contribution of SMR was negligible [19]. Hence, it is necessary to determine the primary contribution of the yz-plane to the magnetoresistance in all-metallic heterostructures and re-estimate the magnitude of SMR. For this purpose, we studied the Pt/(CoNi)$_n$ multilayer system, which is a magnetic heterostructure with a tunable perpendicular magnetic anisotropy (PMA), high spin polarization, and low Gilbert damping [20-22]; this system has been studied extensively in the field of spintronics. We found that the magnitude of SMR in this structure is significantly lower than the theoretically predicted value.

## II. THIN-FILM DEPOSITION AND STRUCTURAL CHARACTERIZATION

In this work, Ta (3)/Pt (10)/[Co (0.3) Ni (0.4)]$_n$/Ta (3) was deposited on a naturally oxidized monocrystalline silicon substrate via magnetron sputtering at room temperature. The numbers in brackets represent the thickness of each layer in nanometer units, and $n$=2, 3, 4, 5, 6, 7, 8, 9, 11, and 12, which are the number of repetitions of the magnetic double-layer. Ta in the upper layers forms a buffer, whereas that in the lower layer forms protective layers; this reduces the influence of substrate roughness and prevents oxidation of the film. During sputtering, the base vacuum is lower than $5\times10^{-8}$ Torr, and the Ar pressure is maintained at $3\times10^{-3}$ Torr. The static magnetic properties of all the samples were determined using vibrating-sample magnetometry, and the crystal structures were characterized via high-resolution x-ray diffractometry (HR-XRD) measurements. Furthermore, magnetoresistance was characterized using the physical property measurement system (PPMS; Quantum Design in SandieDo, USA), and a self-built system was employed for the ST-FMR test.

First, we determined the crystal structure and orientation of all samples by using HR-XRD with a Cu $K\alpha$ radiation source. Figure 1(a) presents the $\theta-2\theta$ x-ray diffraction spectra of the samples with different repetition numbers; in this figure, the dashed lines at 39.7° and 44.5° represent the expected positions of the Pt (111) and CoNi (111) peaks, respectively, in unstrained lattices [23]. For the $n$=2 sample, only the Pt (111) peak is observed in the range of 34°–50°. Several clear secondary diffraction peaks generated by Laue oscillations are noted around this peak, indicating that Pt has high crystallinity in the (111) orientation and low interface roughness. When $n > 4$, a relatively weak peak appears at 42°–44°. With an increase in $n$, the strength of this peak gradually increases, and the peak position approaches 44.5°. The full width at half maximum of the rocking curves around the Pt (111) and CoNi (111) reflections are determined as 5.7° and 4.9°, respectively, using Gaussian fitting, as shown in Fig. 1(b–c). This indicates that the Pt and CoNi in the samples have (111) textures parallel to the normal direction of the film. To verify that the number of repetitions in this series of samples was precisely controlled, the dependence of the ratio of the intensity of CoNi (111) and Pt (111) on $n$ was analyzed, as shown in Fig. 1(d). From this figure, it is evident that a straight line passing through the origin coincides with the experimental results (denoted as open circles). Therefore, assuming that the thickness of Pt remains unchanged, altering the number of repetitions increases the number of crystal surfaces participating in Bragg reflection; however, it does not affect the crystal structure of each CoNi layer.

To further validate that the CoNi repetition was accurately regulated, we evaluated the dependence of PMA on $n$. As shown in Fig. 2 (a) and (b), an abnormal Hall resistance curve was observed for $n=8$ when a magnetic field was applied along the normal direction of the film and parallel to the film. It can be deduced that the square-shaped anomalous Hall effect (AHE), as shown in Fig. 2(a), exhibits PMA [24]. Based on the field-in-plane loop, the effective perpendicular anisotropy field ($H_k^{\text{eff}}$) can be estimated to be approximately 10 kOe. To determine the precise value of $H_k^{\text{eff}}$, we use the following formula:

$$\frac{\rho_{xy}(H_x)}{\rho_{xy}(0)} = \sqrt{1 - \left(\frac{H_x}{H_k^{\text{eff}}}\right)^2} \quad (1)$$

Thus, the AHE curve was fitted in the range of 1–6 kOe, as shown in the inset of Fig. 2(b) [25]. The fitting result of $H_k^{\text{eff}}$ is $7.6 \pm 0.1$ kOe. Thereafter, we can obtain the relationship between $H_k^{\text{eff}}$ and $n$ according to this fitting method, as shown in Fig. 2(c). The solid line depicted in this figure represents the fitting curve of $A/n+B$; this line proves that the PMA can be primarily attributed to the interface effect. This result is consistent with previous reports [23]. Using the analytical method proposed by You et al. [26] and neglecting the volume anisotropy energies of Co and Ni, the average perpendicular anisotropic energy per unit area for each CoNi layer can be expressed as

$$K_{\text{eff}}D + 2\pi DM_s^2 = K_s^{\text{Co/Ni}} + K_s^{\text{Ni/Co}} + (1/n)(K_s^{\text{Pt/Co}} + K_s^{\text{Ni/Ta}} - K_s^{\text{Ni/Co}}) \quad (2)$$

In this equation, the effective perpendicular anisotropy can be obtained using $K_{\text{eff}} = \frac{H_k^{\text{eff}} M_s}{2}$; $D=0.7$ nm is the bilayer thickness; $K_s^{\text{Co/Ni}}$, $K_s^{\text{Ni/Co}}$, $K_s^{\text{Pt/Co}}$, $K_s^{\text{Ni/Ta}}$ are the interface anisotropy energies of the Co/Ni, Ni/Co, Pt/Co, and Ni/Ta interfaces, respectively; and the second term on the left-hand side is the demagnetizing energy, where $M_s = (M_s^{\text{Co}} t_{\text{Co}} + M_s^{\text{Ni}} t_{\text{Ni}})/D = 614.3$ emu·cm$^{-3}$. As shown in Fig. 2(d), there exists a linear relationship between this energy and $1/n$. Considering $K_s^{\text{Co/Ni}} = K_s^{\text{Ni/Co}}$ and neglecting the term $K_s^{\text{Ni/Ta}}$, $K_s^{\text{Co/Ni}} = 0.14 \pm 0.01$ (erg·cm$^{-2}$) and $K_s^{\text{Pt/Co}} =$

$0.57 \pm 0.04$ (erg·cm$^{-2}$) are obtained from Eq. (2). These values are similar to those reported by previous studies [26, 27]. Thus, based on the abovementioned analysis of interface anisotropy, it can be concluded that the samples in this study exhibit clear periodic structures and Co/Pt and Co/Ni interfaces.

### III. ANGLE DEPENDENCE OF MAGNETORESISTANCE

It is generally believed that the presence of cos^2 symmetric magnetoresistance in the HM/FM structure in the plane perpendicular to the direction of current [i.e., the yz-plane, as defined in Fig. 3(a)] is a sign of the SMR effect [1]. However, in order to strictly separate the contributions of SMR from the geometrical-size effect (GSE) [28, 29] and anisotropic interface magnetoresistance (AIMR) [16, 17], it is necessary to determine the dependence of magnetoresistance on thickness for the xy- and yz-planes; the results for both these planes need to be analyzed simultaneously.

As shown in Fig. 3, we measure the angle-dependent magnetoresistance in the xy- and yz-planes. A magnetic field of 9 T is applied during the experiment to ensure that the magnetic moment of the FM is parallel to the magnetic field. α and β denote the angles of the magnetic field direction measured from the *x*-axis in the xy-plane and the *z*-axis in the yz-plane, respectively. The experimental results for three typical samples are shown in Fig. 3(b). As shown in this figure, the $\cos^2\alpha$ behavior is observed for all CoNi thicknesses (denoted by black circles), which is consistent with the behavior of conventional AMR. However, the magnetoresistance curve in the yz-plane includes a higher-order cosine term, and the contribution of this higher-order term increases with *n*. The polar diagram can be used to further highlight the symmetry of magnetoresistance in the yz-plane with the changes in *n*, as shown in Fig. 3(d). Subsequently, the magnetoresistance curve for the xy-plane is fitted using the conventional AMR equation [30]:

$$\rho(\alpha) = \rho_0 + \Delta\rho_{xy}\cos^2\alpha \qquad (3)$$

where $\rho_0$ refers to the resistivity of the sample when the magnetic field direction is perpendicular to the direction of current, and $\Delta\rho_{xy}$ refers to the change in resistivity when the magnetic field coincides with the *x*- and *y*-axes, respectively. Results of the Fourier analysis reveal that the orders of $\cos^{2n}\beta$ need to be considered in order to appropriately describe the magnetoresistance curve for the yz-plane [19], according to the following equation :

$$\rho(\beta) = \rho_0 + \sum_{n=1}^{3} \Delta\rho_{yz}^{(2n)} \cos^{2n}\beta \qquad (4)$$

This adequately fits the experimental data, as shown in Fig. 3(c). Thereafter, we obtain the contribution of the second-order ($\cos^2\beta$), fourth-order ($\cos^4\beta$), and sixth-order ($\cos^6\beta$) terms based on the magnetoresistance of the yz-plane. For all the samples, the sixth order is considerably smaller than the other two terms; therefore, it is omitted from subsequent discussions.

It can be concluded that all the magnetoresistance effects in the sample are caused by a single magnetic layer, and the relationship between magnetoresistance and the thickness of the magnetic layer ($t_{\text{CoNi}}$) can be analyzed [16, 17, 19]. In this model, the

SMR effect, which should have appeared in the Pt layer, is ascribed to an interfacial magnetoresistance effect that is inversely proportional to the FM thickness. Based on this approach, we first consider the shunt effect, as follows:

$$\frac{\Delta\rho_{CoNi}}{\rho_{CoNi}} = \frac{\Delta\rho}{\rho} \times \frac{d_{Pt}\rho_{CoNi}+t_{CoNi}\rho_{Pt}}{t_{CoNi}\rho_{Pt}} \quad (5)$$

where $\frac{\Delta\rho}{\rho}$ is the total magnetoresistance observed during the experiment, and $\rho_{CoNi}=1.8\times10^{-7}$ Ω·m and $\rho_{Pt}=2\times10^{-7}$ Ω·m represent the resistivities of CoNi and Pt, respectively. In addition, considering the presence of magnetic dead layers at the interface between CoNi and the non-magnetic layer, we employed the effective thickness of the magnetic layer ($t_{CoNi,corr} = t_{CoNi} - 0.65$ nm), as shown in Fig. 4.

The SMR effect is derived through the absorption and reflection of the interface spin current in the Pt layer, which is related to the direction of FM magnetization and is reflected as magnetoresistance under the combined effects of spin Hall effect (SHE) and inverse SHE. There are three basic characteristics of this effect. First, the variation in magnetoresistance depends on the angle between magnetization and spin polarization ($y$-axis); hence, it can be observed in both the yz- and xy-planes. Second, according to a previous theory, there is only one cos^2 term [1-3]. Finally, when SMR is caused by the magnetoresistance of the CoNi layer, it exhibits an inverse relationship with the thickness of the CoNi layer. As shown in Fig. 4(a), the magnetoresistance ratio in the xy-plane ($\Delta\rho_{xy,CoNi}/\rho_{CoNi}$) is linearly dependent on the thickness of $t_{CoNi,corr}$ and is not inversely proportional to the thickness as expected. This result indicates that the samples may not possess SMR. Nevertheless, magnetoresistance is observed in the yz-plane, which is typically the basis for identifying the presence of SMR. Therefore, the magnetoresistance observed in the yz-plane needs to be further analyzed.

Figure 4(b) indicates a non-monotonic change between the magnetoresistance ratios in the yz-plane ($\Delta\rho^{(2)}_{yz,CoNi}/\rho_{CoNi}$) and $t_{CoNi,corr}$. When $t_{CoNi,corr}<4$ nm, the magnetoresistance ratio decreases with the increase in thickness, thereby confirming the existence of an interface contribution. When $t_{CoNi,corr} \geqslant 4$ nm, the magnetoresistance ratio tends to increase with the thickness. Therefore, we analyze and fit the dependence of the magnetoresistance ratio on thickness, using the following empirical formula:

$$\frac{\Delta\rho^{(2n)}_{yz,CoNi}}{\rho_{CoNi}} = \frac{A}{t_{CoNi}} + B \cdot t_{CoNi} + C \quad (6)$$

This formula introduces three undetermined parameters: *A*, *B*, and *C*. *A* describes the contribution of the interface, *B* represents the contribution of the linear relationship with thickness, and *C* represents the contribution of magnetoresistance independent of thickness. As indicated by the black curve in Fig. 4(b), the fitting formula is in good agreement with the experimental data; the fitting results are *A* = 0.025 nm, *B* = 0.012 nm$^{-1}$, and *C* = -0.0006. Assuming that *A* entirely comprises SMR contributions, a small percent change in magnetoresistance should have been observed. However, as shown in Fig. 4(a), when the thickness is less than 2 nm, the magnetoresistance is below 1%, which further proves that the contribution of SMR in *A* can be neglected.

Three primary factors contribute toward the magnetoresistance in the yz-plane. First, AIMR arises because of the anisotropic interfacial scattering of conductive electrons, which mainly presents when magnetization in the yz-plane varies with high-order cosine terms, as indicated by the red dashed line in Fig. 4(b). Second, the negative constant term in the second-order term can be attributed to the GSE effect, which is caused by crystallinity and consequently the anisotropic orientation of grains [16, 19]. This is associated with *C* in the fitting formula. Third, the part where there is a linear relationship with the thickness appears in the second- and fourth-order terms; the case of this phenomenon is currently under investigation. In summary, based on the abovementioned detailed analyses, we proved that SMR was not evident in the samples. Moreover, we also reveal that the presence of the cos^2 symmetric magnetoresistance in the yz-plane is not the only criterion for determining the existence of SMR.

## IV. THEORETICAL PREDICTION OF SMR AMPLITUDE

To further investigate this absence of SMR, based on the analysis results depicted in Fig. 4, the theoretical value of SMR in the sample needs to be determined. When the Pt layer thickness is significantly larger than its spin diffusion length, the spin Hall angle ($\theta_{SH}$) is the only parameter that influences SMR [5]. Therefore, using the ST-FMR method, we introduce a microwave signal with a GHz frequency in the Pt/(CoNi)$_n$ microstrip. The typical measured ST-FMR spectrum is shown in Fig. 5(a); the experimental conditions are 18.5 GHz, 25 dBm, and $\alpha=45°$. The spectrum can be well fitted using a general line-shape equation, as follows:

$$V_{dc} = U_s \frac{\Delta H^2}{\Delta H^2 + (H-H_0)^2} + U_a \frac{\Delta H(H-H_0)}{\Delta H^2 + (H-H_0)^2} \quad (7)$$

where $U_s$ and $U_a$ are the voltage amplitudes of the symmetric Lorentz and antisymmetric dispersive line-shapes, respectively; Sankey et al. reported that these line-shapes are a result of the damping-like torque caused by the SHE of the Pt layer and the torque generated by the Oersted field [31]. $\Delta H$ and $H_0$ are the linewidth and resonance field, respectively. To accurately determine the spin Hall angle, we first obtained the angular dependence of $U_s$ and $U_a$, as shown in Fig. 5(b). These dependences satisfy the theoretical angle dependence relationship of $\sin2\alpha\cos\alpha$, which implies that the ratio between $U_s$ and $U_a$ is independent of the angle. In addition, we measured the $U_s/U_a$ ratio of samples with different CoNi thicknesses at 18 GHz and fitted them according to Ref. [32]:

$$\frac{U_s}{U_a}[1 + \left(\frac{4\pi M_{eff}}{H_0}\right)]^{\frac{1}{2}} = \theta_{SH} \frac{\hbar}{e\mu_0 M_s t_{CoNi} d_{Pt}} \quad (8)$$

Based on the reciprocal relationship with $t_{CoNi}$, the spin Hall angle of Pt in the sample was determined to be 0.07. It should be noted that the interface was considered to be transparent to the spin current generated in Pt. If the transparency of the Pt/Co interface T=0.65±0.06 is considered, according to Ref. [33], the spin current density of Pt flowing to the Co/Pt interface is greater than that flowing to the CoNi layer. Thus, the spin Hall angle in our system is estimated to be larger than 0.07. In addition, SMR was

estimated to be larger than 7.3×10⁻⁴ using the following equation:

$$\frac{\Delta \rho_{SMR}}{\rho_{Pt}} \sim \theta_{SH}^2 \frac{\lambda_{Pt}}{d_{Pt}} \tanh\left(\frac{d_{Pt}}{2\lambda_{Pt}}\right) \left[1 - \frac{1}{\cosh\left(\frac{d_{Pt}}{\lambda_{Pt}}\right)}\right] \tag{9}$$

where $\lambda_{Pt}$=1.5 nm is the spin diffusion length of Pt [3].

To directly compare the theoretical value of SMR with the experimental results, the theoretical SMR value is attributed to the magnetoresistance of the CoNi layer through shunt treatment; this resulted in a contribution that is inversely proportional to the thickness of the CoNi layer in the xy- and yz-planes, corresponding to $A_{SMR} \approx 6.57 \times 10^{-3}$ nm in Eq. (6). To facilitate comparison, the theoretical prediction of SMR, second-order contribution of AIMR in experimental data, and experimental results of magnetoresistance in the xy-plane are all included in Fig. 5(d). It is evident that the theoretical SMR is higher than the experimental value, when the thickness of the CoNi layer is less than 2 nm. This indicates that the accuracy of these experimental results can fully characterize the SMR effect predicted theoretically. However, the results did not show a contribution in inverse proportion to the thickness. Further estimates based on our experimental accuracy suggest that the order of magnitude of SMR in our system should be less than 1×10⁻⁴. This value is considerably lower than previously reported SMR values for metallic systems with Pt = 10 nm [4, 12]. In addition, the magnitude of AIMR is 4 times greater than that of SMR. Therefore, although SMR was noted, the magnetoresistance in the yz-plane is dominated by the contribution of AIMR.

## V. DISCUSSION

The results of the ST-FMR experiment indicate that the spin current torque generated in the Pt layer affects the magnetization of CoNi,; however, the absorption of these spin currents was not evidenced by an apparent SMR effect. Based on the spin diffusion equation, the spin current density $j_s^{(F)}$ at the magnetic interface is [1]

$$e j_s^{(F)}(M) = G_r M \times (M \times \mu_s) + G_i (M \times \mu_s) \tag{10}$$

where $G_r(G_i)$ is the real(imaginary) component of the spin mixing conductance, and $\mu_s$ is the accumulation of spin at the interface. The spin current flowing into the FM through the interface and the reflected spin current are uniquely influenced by the direction of magnetization, which eventually leads to the SMR effect. Therefore, starting from the interface effect, we consider the effect of spin memory loss (SML) on the spin transport of the Pt/Co interface [34, 35]. SML implies that the spin current flowing into the CoNi layer is significantly lower than that flowing into the Pt/Co interface in Pt. This type of interfacial absorption of the spin current is independent of the **M** of the CoNi layer. However, SML is equivalent to inserting a spin sink layer (SSL) with a specific thickness and spin diffusion length between the FM and the HM. Therefore, the spin boundary conditions between SSL and FM are similar to Eq. (10), and the Pt layer and SSL can be combined to form an equivalent HM with a smaller $\theta_{SH}$. As the spin current acting on **M** in the FM can be detected using ST-FMR, $\theta_{SH} = 0.07$, which neglects the interface effect, is lower than previously reported experimental values [33]. However, the magnitude of SMR estimated using $\theta_{\text{SH}} = 0.07$ is still within the range of our experimental accuracy. Therefore, only considering the SML does not reasonably explain our experimental results.

As SMR was first confirmed and systematically studied in HM/YIG, we believe that the SMR effect in HM/FMI and HM/FM is significantly different due to the different carriers of spin current and the corresponding scattering mechanisms in FMI and FMs. In general, spin current can only be absorbed by FMI through the spin-angular-momentum exchange between the localized magnetization **M** in FMI and the conduction-electron spin polarization **σ** in HMs, i.e., the spin-orbit torque (SOT). However, in FMs, the spin relaxation of itinerant electrons needs to be considered. In 2011, Berger extended Elliott's theory of spin relaxation in metals and semiconductors to include metallic ferromagnets [36]. He found that the spin relaxation in FMs arises from the spin-orbital interaction associated with the crystalline periodical potential and the random potential caused by scatters; this is proportional to the spin down resistivity for materials whose spin-up fermi levels are located above the top of the 3$d$ band, such as Ni and Co. Further experimental results confirmed this theory; for instance, Sagasta et.al. reported that the Elliott–Yafet mechanism is the dominant spin relaxation mechanism in permalloys [37].

With regard to the absence of SMR observed during our experiment, we consider two mechanisms that absorbed the injected spin current in FMs: SOT (**M**-dependent, interfacial effect) and spin relaxation (less **M**-independent, bulk effect). The absorption of spin currents using SOT is dominant when **M** is perpendicular to **σ**. Contrarily, when **M** is parallel to **σ**, SOT is not applicable; however, most of the spin currents would be relaxed in FMs. In general, most of the spin currents flowing into FMs will be absorbed, resulting in a spin boundary condition for the FM/HM, which is independent of (or less dependent on) the direction of magnetization. Therefore, although SOT in the Pt/CoNi system was evident, a corresponding magnitude of SMR does not appears. In conclusion, we believe that, when considering different mechanisms for the spin current absorption in FMs and FMI, boundary conditions of the spin current in the HM/FMI may not be applicable to the HM/FM interface; thus, SMR performance would be significantly different.

## VI. CONCLUSIONS

In summary, we systematically investigated the magnetoresistance effect in a Pt/(CoNi)$_n$ system with strong PMA and a (111) texture. By ascribing all the magnetoresistance to the CoNi layer, the contribution of the magnetoresistance with interface characteristics was emphasized. By comparing the magnetoresistance characteristics of the yz and xy planes, we determined that the magnetoresistance with interface characteristics in the yz plane could be primarily attributed to AIMR, and no obvious contribution of SMR was found in the experimental results. However, we obtained the $\theta_{\text{SH}}$ of Pt through an ST-FMR experiment, and the magnitude predicted by the SMR theory was higher than our experimental accuracy. Consequently, the appearance of $\cos^2\beta$ terms in the plane perpendicular to the current cannot be used as an indicator of SMR in metallic magnetic heterostructures. The separation of the magnetoresistance in the yz plane introduced by the symmetry breaking of the interface is a necessary prerequisite for the study of the SMR effect in metallic magnetic heterostructures.

**Acknowledgments** This project was supported by NSFC (Nos 11429401, 91963201 and 51471081), the Program for Changjiang Scholars and Innovative Research Team in University (No IRT-16R35), the Creative Project of Key Laboratory for Magnetism and Magnetic Materials of the Ministry of Education, and the 111 Project under Grant No. B20063. We would like to thank Editage (www.editage.cn) for English language editing.

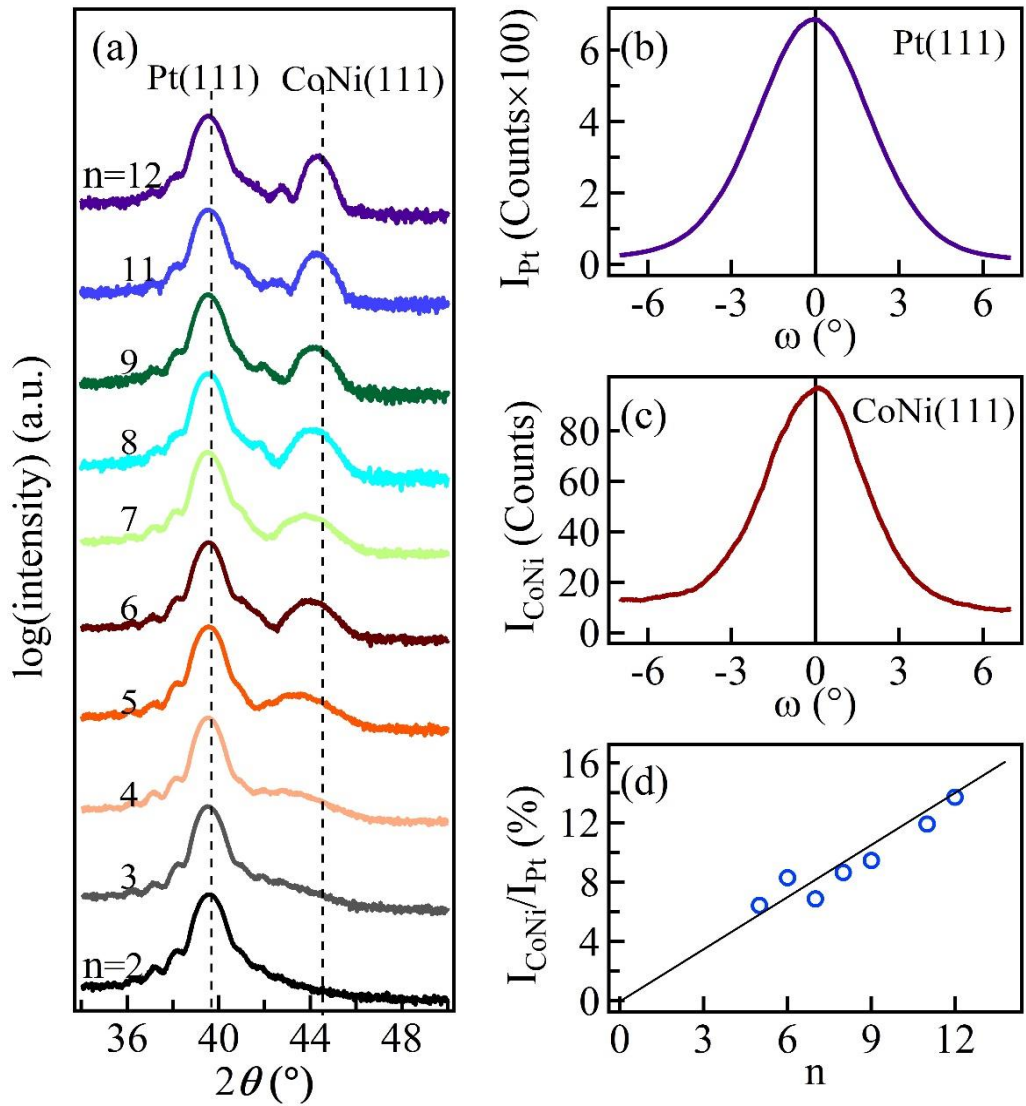

Fig. 1. (a) Out-of-plane x-ray measurement of as-deposited Ta(3)/Pt(10)/[Co(0.3)/Ni(0.4)]n/Ta(3) for n ranging from 2 to 12. Dashed lines at 39.7° and 44.5° represent the expected positions of unstrained Pt(111) and Co/Ni(111) peaks, respectively. (b,c) Rocking curves around Pt (111) and CoNi (111) with a reflection of n=12. (d) Blue data points indicate the dependence of the peak intensity ratio of CoNi and Pt on the number of repetitions n; as indicated, the straight black line, which was fitted to the experimental results, passes through the origin.

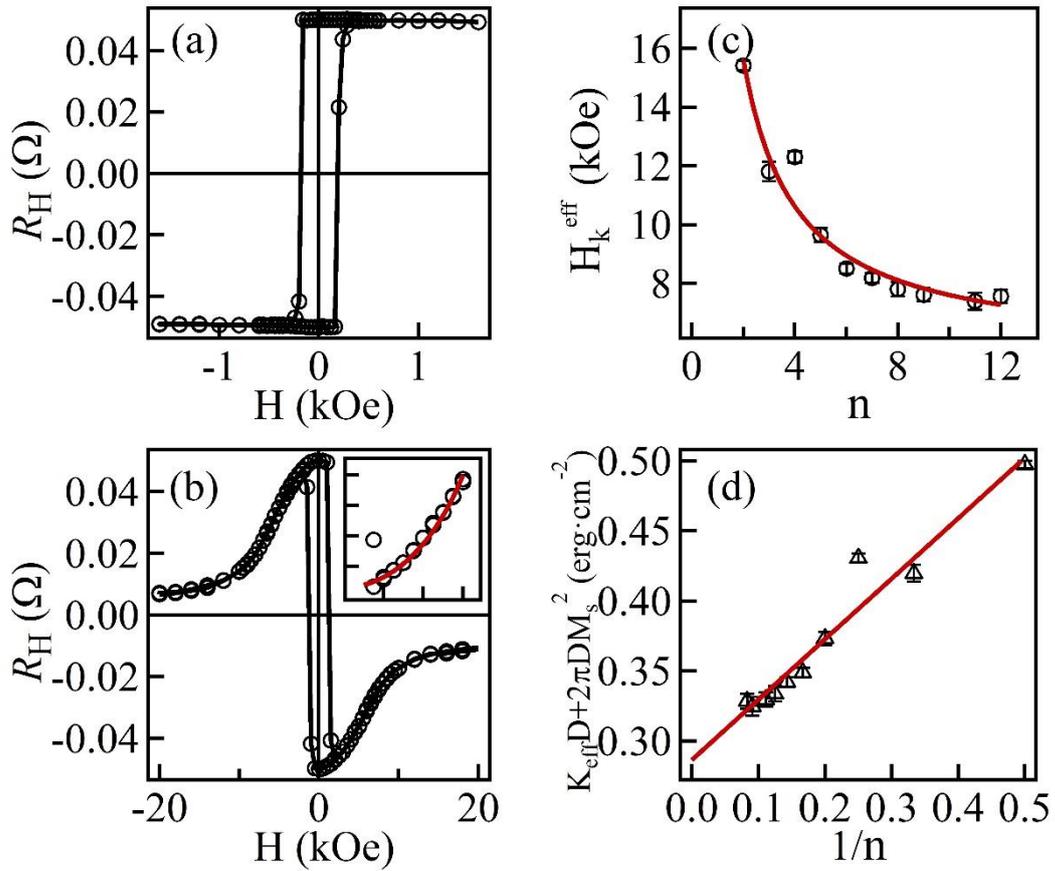

Fig. 2 (a,b) Loops of Hall resistance when a magnetic field was applied along the film normal and plane of $n$=8. (c) Black circles represent the dependence of the effective perpendicular anisotropy field on the repetition number, which is well fitted by $A/n+B$ ($A$ and $B$ are constants), as indicated by the curve. (d) Triangle represents the perpendicular anisotropy energy per unit area, and the red line denotes the linear fitting. The inset in (b) depicts local enlarged data from 1–6 kOe, and the curve is the result of fitting using Eq. (1).

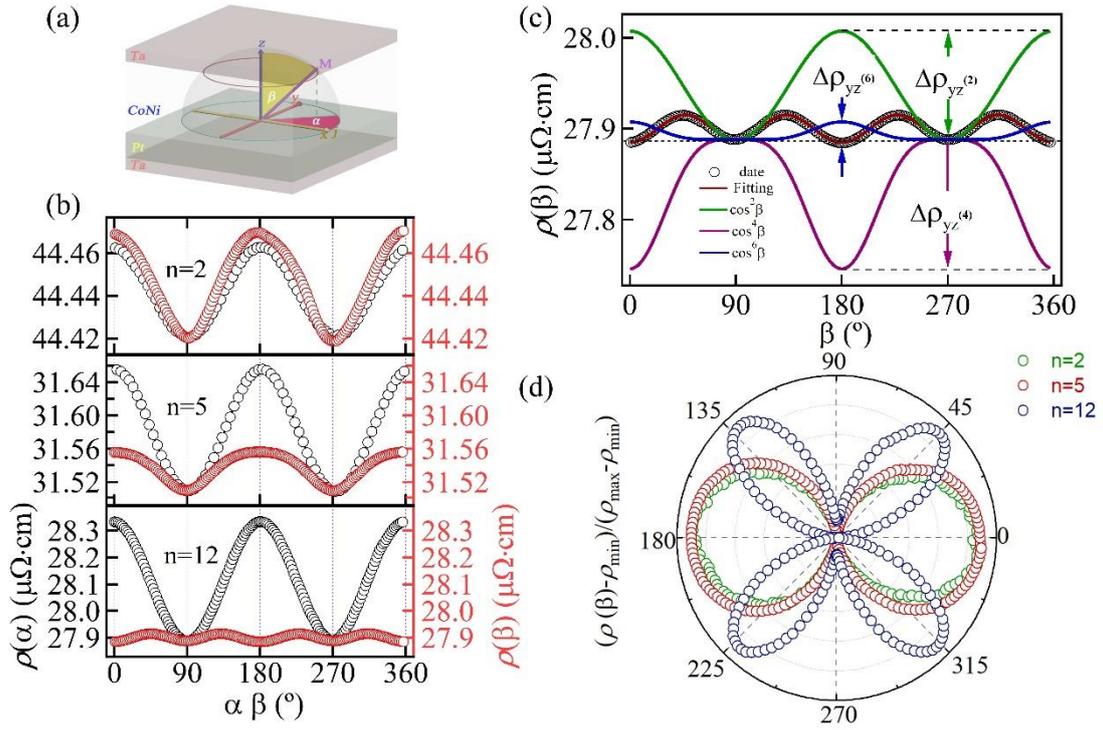

Fig. 3. MR behavior of Pt/[Co/Ni]$_n$ structures. (a) Illustration of the Ta/Pt/[Co/Ni]$_n$/Ta sample structure and coordinate system of the experiment. (b) Resistivity ρ with respect to the magnetization orientation for three samples with different repetitions. The rotation of magnetization **M** is performed in the film plane (β=90°, denoted as black dots) and in the plane perpendicular to current **j** (α=90°, denoted as red dots). ρ(α, β=90°) exhibits the conventional AMR behavior for all samples (cos$^2$α fits), and the functional dependence of ρ(α=90°, β) varies with n. (c) Individual cos$^{2n}$β contributions to ρ(α=90°, β) for n=12. (d) Polar plot of ρ(α=90°, β) for the three samples.

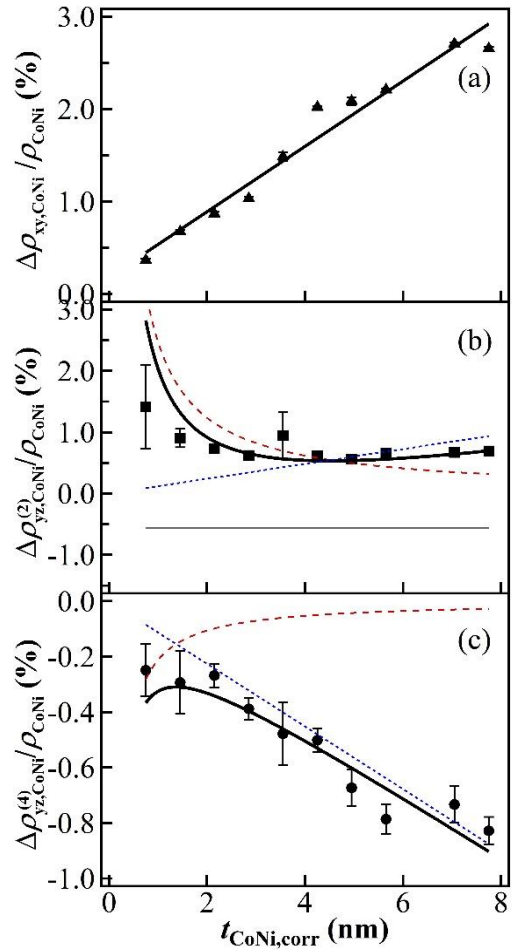

Fig. 4. Thickness-dependent magnetoresistance after corrections. (a) $\Delta\rho_{xy,CoNi}/\rho_{CoNi}$, (b) $\Delta\rho^{(2)}_{yz,CoNi}/\rho_{CoNi}$, and (c) $\Delta\rho^{(4)}_{yz,CoNi}/\rho_{CoNi}$ as a function of the thickness of CoNi. The symbols denote experimental data, whereas the solid black lines represent fitting lines. The dashed lines (dotted lines) represent the contribution of AIMR (this contribution increases linearly with thickness). The GSE contribution is a constant that is independent of thickness; it is indicated by a thinner solid straight line in (b).

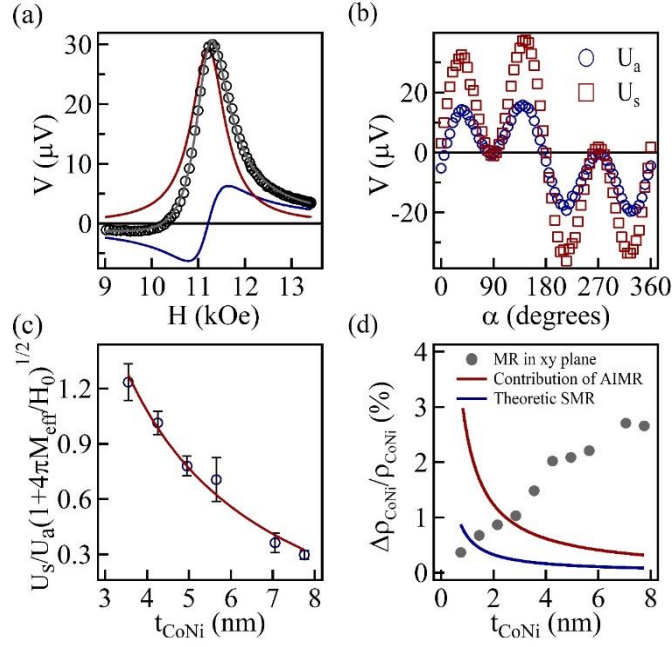

Fig. 5. (a) ST-FMR spectrum of for the n = 7 sample measured at 45° in-plane, presenting antisymmetric (blue curve) and Lorentz symmetric (red curve) line-shapes. (b) Angular dependence of $U_s$ and $U_a$ at the same frequency in the xy-plane. (c) Fitting of the spin Hall angle ($\theta_{SH}$) according to Eq. (8). (d) Dependence of xy-plane magnetoresistance, theoretical value of SMR, and experimental value of AIMR on the thickness of CoNi.